\documentclass[aps,preprint]{revtex4}
\usepackage{graphicx}
\usepackage[scanall]{psfrag}
\usepackage{amssymb}

\begin{document}

%		DEFINITIONS FOR TEX
%
%%%%%%%%%%%%%%%%%%%%%%%%%%%%%%%%%%%%%%%%%%%%%%%%%%%%%%%%%%%%%%%
%
%
%\def\e{\'e}
%\def\ee{\`e}
%%%%%%%%%%%%%%%%%%%DEFINITIONS%%%%%%%%%%%%%%%%%%%%%%%%%%%%%%%%%
%
\def\oti{{\otimes}}
\def\lb{ \left[ }
\def\rb{ \right]  }
\def\tilde{\widetilde}
\def\bar{\overline}
\def\hat{\widehat}
\def\*{\star}
\def\[{\left[}
\def\]{\right]}
\def\({\left(}		\def\BL{\Bigr(}
\def\){\right)}		\def\BR{\Bigr)}
	\def\BBL{\lb}
	\def\BBR{\rb}
%
%%%%%%%%%%%%%%%%%%%%%%%%%%%%%%%%%%%%%%%%%%%%%%%%%%%%%%%%%%%%%%%
%
\def\zb{{\bar{z} }}
\def\zbar{{\bar{z} }}
\def\frac#1#2{{#1 \over #2}}
\def\inv#1{{1 \over #1}}
\def\half{{1 \over 2}}
\def\d{\partial}
\def\der#1{{\partial \over \partial #1}}
\def\dd#1#2{{\partial #1 \over \partial #2}}
\def\vev#1{\langle #1 \rangle}
\def\ket#1{ | #1 \rangle}
\def\rvac{\hbox{$\vert 0\rangle$}}
\def\lvac{\hbox{$\langle 0 \vert $}}
\def\2pi{\hbox{$2\pi i$}}
\def\e#1{{\rm e}^{^{\textstyle #1}}}
\def\grad#1{\,\nabla\!_{{#1}}\,}
\def\dsl{\raise.15ex\hbox{/}\kern-.57em\partial}
\def\Dsl{\,\raise.15ex\hbox{/}\mkern-.13.5mu D}
%
%%%%%%%%%%%%%%%%%%%%GREEK LETTERS%%%%%%%%%%%%%%%%%%%%%%%%%%%%%%
%
%\def\th{\theta}		\def\Th{\Theta}
\def\ga{\gamma}		\def\Ga{\Gamma}
\def\be{\beta}
\def\al{\alpha}
\def\ep{\epsilon}
\def\vep{\varepsilon}
\def\la{\lambda}	\def\La{\Lambda}
\def\de{\delta}		\def\De{\Delta}
\def\om{\omega}		\def\Om{\Omega}
\def\sig{\sigma}	\def\Sig{\Sigma}
\def\vphi{\varphi}

%
%%%%%%%%%%%%%%%%%%%CALIGRAPHIC LETTERS%%%%%%%%%%%%%%%%%%%%%%%%%
%
\def\CA{{\cal A}}	\def\CB{{\cal B}}	\def\CC{{\cal C}}
\def\CD{{\cal D}}	\def\CE{{\cal E}}	\def\CF{{\cal F}}
\def\CG{{\cal G}}	\def\CH{{\cal H}}	\def\CI{{\cal J}}
\def\CJ{{\cal J}}	\def\CK{{\cal K}}	\def\CL{{\cal L}}
\def\CM{{\cal M}}	\def\CN{{\cal N}}	\def\CO{{\cal O}}
\def\CP{{\cal P}}	\def\CQ{{\cal Q}}	\def\CR{{\cal R}}
\def\CS{{\cal S}}	\def\CT{{\cal T}}	\def\CU{{\cal U}}
\def\CV{{\cal V}}	\def\CW{{\cal W}}	\def\CX{{\cal X}}
\def\CY{{\cal Y}}	\def\CZ{{\cal Z}}

\def\rvac{\hbox{$\vert 0\rangle$}}
\def\lvac{\hbox{$\langle 0 \vert $}}
\def\comm#1#2{ \BBL\ #1\ ,\ #2 \BBR }
\def\2pi{\hbox{$2\pi i$}}
\def\e#1{{\rm e}^{^{\textstyle #1}}}
\def\grad#1{\,\nabla\!_{{#1}}\,}
\def\dsl{\raise.15ex\hbox{/}\kern-.57em\partial}
\def\Dsl{\,\raise.15ex\hbox{/}\mkern-.13.5mu D}
%
%%%%%%%%%%%%%%%%%%%%GREEK LETTERS%%%%%%%%%%%%%%%%%%%%%%%%%%%%%%
%
%%%%%%%%%%%%%%% MATH CHARACTERS %%%%%%%%%%%%%%%%%%%%%%%%%%%%
%
\font\numbers=cmss12
%\font\numbers=cmu10 scaled\magstep1
\font\upright=cmu10 scaled\magstep1
\def\stroke{\vrule height8pt width0.4pt depth-0.1pt}
\def\topfleck{\vrule height8pt width0.5pt depth-5.9pt}
\def\botfleck{\vrule height2pt width0.5pt depth0.1pt}
\def\Zmath{\vcenter{\hbox{\numbers\rlap{\rlap{Z}\kern
0.8pt\topfleck}\kern 2.2pt
                   \rlap Z\kern 6pt\botfleck\kern 1pt}}}
\def\Qmath{\vcenter{\hbox{\upright\rlap{\rlap{Q}\kern
                   3.8pt\stroke}\phantom{Q}}}}
\def\Nmath{\vcenter{\hbox{\upright\rlap{I}\kern 1.7pt N}}}
\def\Cmath{\vcenter{\hbox{\upright\rlap{\rlap{C}\kern
                   3.8pt\stroke}\phantom{C}}}}
\def\Rmath{\vcenter{\hbox{\upright\rlap{I}\kern 1.7pt R}}}
\def\Z{\ifmmode\Zmath\else$\Zmath$\fi}
\def\Q{\ifmmode\Qmath\else$\Qmath$\fi}
\def\N{\ifmmode\Nmath\else$\Nmath$\fi}
\def\C{\ifmmode\Cmath\else$\Cmath$\fi}
\def\R{\ifmmode\Rmath\else$\Rmath$\fi}
%%%%%%%%%%%%%%%%%%%%%%%%%%%%%%%%%%%%%%%%%%%%%%%%%%%%%%%%%%%%%%%%%
 %%%%%%%%%%%%%%%%%% END OF DEFINITIONS %%%%%%%%%%%%%%%%%%%%%%
 %%%%%%%%%%%%%%%%%%%%%%%%%%%%%%%%%%%%%%%%%%%%%%%%%

\def\barray{\begin{eqnarray}}
\def\earray{\end{eqnarray}}
\def\beq{\begin{equation}}
\def\eeq{\end{equation}}

\def\no{\noindent}

\def\pmb#1{\setbox0=\hbox{#1}%
\kern-.025em\copy0\kern-\wd0
\kern.05em\copy0\kern-\wd0
\kern-.025em\raise.0433em\box0 }

\def\smallsqrtk{{\scriptstyle \sqrt{k}}}
\def\smallsqrttwo{{ \scriptstyle \sqrt{2}}}
\def\texthalf{{\textstyle \inv{2}}}

\def\glk{{gl(1|1)_k}}
\def\glhat{gl(1|1)}
\def\psibar{\bar{\psi}}
\def\betabar{\bar{\beta}}
\def\ospk{osp(2|2)_k }
\def\osphat{osp(2|2)}
\def\Shat{\hat{S}}

\def\bfphi{\pmb{$\phi$}}
\def\bfchi{\pmb{$\chi$}}
\def\phibar{\bar{\phi}}
\def\chibar{\bar{\chi}}
\def\sqrtk{\sqrt{k}}
\def\rep{{\it r}}
\def\repbar{ {\bar{\it r}} }

\def\glrep#1#2{{\langle #1, #2 \rangle}}
\def\glrepprime#1{{ \langle #1, #1 \rangle'}}
\def\hone#1{{\langle #1\rangle_{(1)}}}
\def\hfour#1{{\langle #1 \rangle_{(4)}}}
\def\Deltahj#1#2{{\Delta_{\glrep{#1}{#2}}}}
\def\chitilde{{\tilde{\chi}}}

\def\Deltachi{ \Delta^{(\chi)}}

\def\Vbos#1#2{\CV^\phi_{#1,#2}}
\def\V#1#2{V_{\glrep{#1}{#2}}}
\def\Vfourh{V_\hfour{h}}
\def\Vfour#1{V_\hfour{#1}}

\def\muhalf{\mu_{1/2}}
\def\sigmahalf{ \sigma_{1/2}}

\def\Ncopy{{\rm N-copy}}
\def\suNhat#1{su(N)_{#1}}
\def\free{{\rm free}}

\def\Deltamin{\Delta^{\rm (min)}}

\def\bfPhi{{\bf \Phi}}
\def\bfVbos#1#2{{\bf V}^\phi_{{#1},{#2}}}
\def\bfell{{\pmb{$\ell$}}}
\def\Hbar{\bar{H}}
\def\Jbar{\bar{J}}
\def\repbar{{\bar{\rep}}}
\def\tbar{\bar{t}}
\def\Vbar#1#2{\bar{V}_{\glrep{#1}{#2}}}
\def\bfmu{\pmb{$\mu$}}
\def\bfsigma{\pmb{$\sigma$}}
\def\bfPhihj#1#2{\bfPhi_{\glrep{#1}{#2}}}

\def\qvec{\vec{q}}
\def\phivec{\vec{\phi}}

\def\osprep#1#2{ [ #1, #2 ]^{osp}}
\def\surep#1#2{ [ #1, #2]^{su}}

\def\sqrttwo{\sqrt{2}}

\def\eight{[8]^{osp}}

\def\smallhalf{{\textstyle \inv{2}}}

\title{The $gl(1|1)$  super-current algebra: the r\^ole of twist
and logarithmic fields}
\author{Andr\'e  LeClair}
\affiliation{Newman Laboratory, Cornell University, Ithaca, NY.
\\ and \\
Isaac Newton Institute for Mathematical Sciences,  Cambridge, England} 
\date{October 2007}

\bigskip\bigskip\bigskip\bigskip

\begin{abstract}

A  free field representation of the $gl(1|1)_k$ current
algebra at arbitrary level $k$ is given in terms of
two scalar fields and a symplectic fermion.    The primary
fields for all representations are explicitly constructed
using the twist and logarithmic fields in the symplectic 
fermion sector.    A closed operator algebra is described
at integer level $k$.    Using a new super spin charge
separation involving $gl(1|1)_N$ and $su(N)_0$,  we 
describe how the $gl(1|1)_N$ current algebra can describe
a non-trivial critical point of disordered Dirac fermions.  
Local $gl(1|1)$ invariant lagrangians are defined which
generalize the Liouville and sine-Gordon theories.   
We apply these new tools to the spin quantum Hall transition
and show that it can be described as a logarithmic perturbation 
of the $osp(2|2)_{k}$ current algebra at $k=-2$.

\end{abstract}

\maketitle

\section{Introduction}

A variety of $2D$ models with Lie super-group symmetry 
are now understood to be important in many diverse
areas of modern theoretical physics.    A partial list  includes
applications to disordered 
systems\cite{Mudry1,Bernard1,Serban,Guruswamy,SpinCharge,Tsvelik2},
statistical mechanics\cite{ReadSaleur,EsslerSaleur},
and string theory\cite{String1,String2,Beisert}.    
Although many results are known,  much remains to be understood
about these models in comparison with their ordinary bosonic
counterparts.

The simplest Lie super-algebra is $gl(1|1)$ and it's current
algebra,  $gl(1|1)_k$ at level $k$,  is the main
subject of this paper.  We also obtain some results for
the $osp(2|2)_{k=-2}$ case. ($osp(2|2)$  is
also referred to as $su(2|1)$ in the literature.)  
The WZNW  sigma-model based on the
$GL(1|1)$ supergroup was considered long ago by Rozanski and
Saleur\cite{RozSal}, and more recently by Schomerus and
Saleur\cite{Schomerus1}  using harmonic analysis on the supergroup.  
The latter analysis was extended to other supergroups 
in\cite{Gotz,Quella}.    In contrast, in our work  the starting point is not
 the WZNW model field theory, but rather 
 the quantum field
theory is  constructed algebraically using the current algebra itself, as was
done for the $su(2)$ theory by Knizhnik and Zamolodchikov\cite{KZ}.  
In this way  new results concerning the spectrum of fields 
are obtained, and explicit constructions of the vertex operators
for all representations are given in terms of twist 
and logarithmic fields.

For the remainder of this introduction, we summarize our main results
and describe the organization of the remainder of the paper. 
After reviewing the definitions of the super current-algebras in section 
II,  we construct a free field representation  in section III
involving  
two scalar fields and a symplectic fermion.   It is known 
from the work\cite{Guruswamy}  that such a representation
exists at level 1,  but it was not evident that this extends
to arbitrary level $k$ with the {\it same field content.}  
In section IV the finite dimensional representations of 
$gl(1|1)$ are reviewed. 
Explicit constructions of the vertex operators are  given in
section VI and  require the twist fields of the symplectic
fermion sector. This r\^ole of twist fields was previously
recognized for the special case of $osp(2|2)_{-2}$ 
by Ludwig\cite{LudwigTwist}.   These twist fields were first studied 
by Kausch\cite{Kausch} and their properties are 
summarized in section V.   Some additional properties of the
twist fields that were needed  are derived in Appendix B.
The vertex operator construction indicates that the level
can be interpreted as a radius of compactification $R=\sqrt{k}$. 
   
The vertex operators for the so-called atypical indecomposable
representations are also explicitly constructed and are logarithmic.
We wish to emphasize that this is only possible in the second-order
description of symplectic fermions because of the  additional zero
modes that are not present in the first-order description.
  
The properties of the twist fields place restrictions on
the allowed spectrum of primary fields and this shows
how to obtain a closed operator algebra (section VII).  
We compare the $k=2$ case with $c=0$ minimal Virasoro models
and thereby show that it is very closely related, but
not identical,  to percolation.

We consider $N$-copies in section VIII and present
a super version of the ordinary spin-charge separation.  
More specifically,  the stress tensor of $N$ free Dirac
fermions and ghosts can be decomposed as the sum of two commuting
pieces which are the stress tensors for $gl(1|1)_{k=N}$ and
$su(N)_{k=0}$.   This generalizes the result 
found in \cite{SpinCharge} for  $N=2$ to arbitrary $N$.
This fact opens up possibilities for
the interpretation of $gl(1|1)_{k=N}$ as a disordered
critical point,  and we explain one simple scenario. 
Our generalization of spin-charge separation to arbitrary
$N$ differs from the one in \cite{BernardSerban} which involves
$osp(2|2)_{-2N} \otimes sp(2N)_0$, and is more relevant to
generalizations of the  spin quantum Hall transition.  
The two are equivalent at $N=2$ since $sp(2) = su(2)$ and
$osp(2|2)_{-2} = gl(1|1)_2$ (See section XI.)

Local (non-chiral) operators that are $gl(1|1)$ invariant
are constructed in section IX.   For the logarithmic representations
these operators can be expressed explicitly in terms of
the free fields and can be used to define $gl(1|1)$ invariant
lagrangians (section X).  In this way we obtain $gl(1|1)$ invariant
versions of the Liouville and sine-Gordon models.  

The $osp(2|2)_{k}$ current algebra at $k=-2$  is known to describe 
the critical point of Dirac fermions subject to a random
$su(2)$ gauge potential\cite{Tsvelik2,SpinCharge}.   
We extend our results to $osp(2|2)_{-2}$ in section XI.
In addition to recovering the results in\cite{LudwigTwist} 
from the $gl(1|1)$ embedding,  we construct the local field
corresponding to the $8$-dimensional logarithmic representation.  

The application of the tools developed in this paper to 
critical points of disordered Dirac fermions is initiated in
section XII where we revisit the spin quantum Hall transition.  
Based on the renormalization group (RG) analysis studied 
in\cite{SpinCharge,networkRG},  we propose that the additional
kinds of disorder in the network model for the spin quantum Hall 
transition can be accounted for by an additional perturbation
of the current algebra $osp(2|2)_{-2}$ 
by the logarithmic operator in the 8-dimensional indecomposable
representation.   We argue this perturbation does not drive
the theory to a new fixed point but rather gives logarithmic
corrections.  This is consistent with the work of Read and Saleur
which emphasized that the critical point possesses $osp(2|2)$ symmetry
but is not precisely a current algebra.

\section{The $\glk$ and $\ospk$ super-current algebras.}

In this section we define   the super-current algebras
and present their stress tensors.  
There are various conventions in the literature for the level $k$.
Our conventions are natural for applications to disordered
Dirac fermions.  Consider the $2D$ free conformal field theory
for a single component $U(1)$ charged Dirac fermion 
$\psi_\pm $ and its ghost partners  $\beta_\pm $, with action
\beq
\label{2.1}
S = \inv{4\pi} \int d^2 x \( 
\psi_-  \d_\zbar \psi_+ + \psibar_- \d_z \psibar_+  
+ \beta_- \d_\zbar \beta_+  + \betabar_- \d_z \betabar_+ \) 
\eeq
where $z,\zbar$ are euclidean light-cone coordinates, 
$z= (x+ i y)/\sqrt{2}$, $\zbar = z^*$.   The ghost fields
have bosonic statistics and the same conformal dimension
as the fermions: $\Delta (\beta_\pm) = \inv{2}$.  First order
systems of this type were treated in generality in \cite{FMS},
in connection with string world sheet ghosts.   In particular
the fermions have Virasoro central charge $c=1$, whereas
the bosons have $c=-1$, and the total central charge is zero.  
The two-point functions of the left-moving fields are 
\beq
\label{2.2}
\langle \psi_- (z) \psi_+ (w) \rangle = 
\langle \psi_+ (z) \psi_- (w) \rangle =
\langle \beta_+ (z) \beta_- (w) \rangle =
- \langle \beta_- (z) \beta_+ (w) \rangle
= \inv{z-w}
\eeq
and similarly for the right-movers, 
$\langle \psibar_-(\zbar) \psibar_+ (\bar{w}) \rangle =
1/(\zbar - \bar{w} )$, etc.  
In the sequel we will not display the right-moving counterparts
if they are the obvious  duplications of the left.  

Define the currents
\beq
\label{2.3}
H = \psi_+ \psi_- ,  ~~~J= \beta_+ \beta_-, ~~~
S_\pm = \pm \psi_\pm \beta_\mp  \, .
\eeq
$H$ and $J$ are the $U(1)$ currents underwhich $\psi_\pm $ 
and $\beta_\pm $ have charge $\pm 1$ and $\mp 1$ respectively. 
Throughout the sequel we will mainly present our results using
operator product expansions (OPE).  
Using eq. (\ref{2.2}) the currents satisfy the $\glk$ super-current
algebra OPE's at $k=1$:
\barray
\nonumber
H(z) H(0) &\sim& \frac{k}{z^2} ,~~~~~ J(z) J(0) \sim - \frac{k}{z^2}
\\ 
\label{2.4}
H(z) S_\pm (0) &\sim& J(z) S_\pm (0) \sim \pm \inv{z} ~ S_\pm (0) 
\\
\nonumber
S_+ (z) S_- (0) &\sim& \frac{k}{z^2} + \inv{z} ~ (H - J )(0)  
\earray
The above $k$ dependence establishes our convention for
$\glk$ at arbitrary $k$.

For a general current $J^a $, define its modes $J^a_n$ 
as follows: $J^a (z) = \sum_{n\in \Zmath} J^a_n \, z^{-n-1}$.  
The modes satisfy the affine Lie superalgebra:
\barray
\nonumber
[H_n, H_m ] &=& - [J_n, J_m] = k \, n \,  \delta_{m+n,0}    
\\ 
\label{2.5}
[H_n , S^\pm_m ] &=& [J_n , S^\pm_m ] = \pm S^\pm_{n+m} 
\\
\nonumber
\{ S^+_n, S^-_m \} &=& k \, n \, \delta_{n+m,0} + H_{n+m} - J_{n+m} 
\earray
The zero modes $H_0, J_0, S^\pm_0$ satisfy the finite $gl(1|1)$ 
algebra.  

The algebra $\glk$ has an inner automorphism that flips the sign of
the level $k$:
\beq
\label{auto}
H\to J, ~~~J\to H,~~~ S_\pm \to \pm S_\pm , ~~~k \to -k
\eeq
This implies that results for negative $k$ can be deduced 
from the case of positive $k$.  

The only additional currents one can define in this theory 
are:
\beq
\label{2.6}
J_\pm = \beta_\mp^2 , ~~~~~\hat{S}_\pm = \psi_\mp \beta_\mp
\eeq
The complete set of currents satisfy the $\ospk$ algebra at level $k$.
The complete set of relations are presented in Appendix A.  
Rescaling $J_\pm \to 2 \sqrt{2} J_\pm $, $J \to 2J$, one
sees that they 
  together satisfy the $su(2)$
current algebra at level $-k/2$.     
Also, making the  
redefinition $H\to -H, J\to J,  \hat{S}_\pm  \to \pm \hat{S}_\pm$, 
one sees that they also satisfy $\glk$, so that $\ospk$ contains
two non-commuting $\glk$'s.

We will need the Sugawara stress tensor $T(z)$.
 The algebra $gl(1|1)$ has
two independent quadratic casimirs:
\beq
\label{casimirs}
C_2 = J^2 - H^2 + S^+ S^- - S^- S^+ , ~~~~~
C_2' = (J-H)^2 
\eeq
where it is implicit  that the above operators are the zero modes
of the currents.   The stress tensor is fixed by the condition
$T(z) J^a (0) \sim J^a (0) /z^2 $,  which requires it to be built
out of both casimirs\cite{RozSal}:
\beq
\label{glstress}
T(z)  = - \inv{2k} \( J^2 - H^2 + S_+ S_- - S_- S_+ \) 
+ \inv{2k^2} (J-H)^2 
\eeq  
The leading term in the OPE $~T(z) T(0)$ shows that $c=0$.

\section{Free field representation}

In this section we present a free field representation of
$\glk$ for any level $k$.  The free Dirac fermion can be
bosonized with a single scalar field.  The results in \cite{FMS} show that  
the first order bosonic $\beta_\pm$ system
can be represented in terms of a single scalar field for the
$U(1)$ current and another first order fermionic $\eta-\xi$
system.  The latter can be formulated as a second order 
symplectic fermion.  (See Appendix B).   Thus, it is clear
that the $k=1$ representation of the last section 
constructed out of the fields
$\psi_\pm , \beta_\pm$ can be represented with two scalar fields
and a symplectic fermion.  What is not so evident is that this
same field content is sufficient to provide a free field construction 
of $\glk$ at any level.  This is in contrast to $su(2)_k$ 
for example where the higher level case requires 
additional  $Z_k$ parafermions.  
(For a review of $2D$ conformal field theory see\cite{CFT,ginsparg}.)

Introduce two scalar fields $\bfphi^a$  and 
a symplectic fermion $\bfchi^a$, $a=1,2$, with the following
free action:
\beq
\label{3.1}
S = \inv{8\pi} \int d^2 x ~ \sum_{a,b=1}^2 \( 
\eta_{ab} \d_\mu \bfphi^a  \d_\mu \bfphi^b + 
\ep_{ab} \d_\mu \bfchi^a \d_\mu \bfchi^b  \) 
\eeq
where
\beq
\label{3.2}
\eta =  \(\matrix{1&0\cr 0 & -1 \cr}\), ~~~~~~
\ep = \( \matrix{0&1\cr -1& 0\cr} \)
\eeq
and $\d_\mu \d_\mu  = 2  \d_z \d_\zbar$.  The $\bfchi$ fields are
Grassman: $(\bfchi^a )^2 = 0$. 
Note that the metric for the bosonic fields  has indefinite signature.
The equations of motion imply that the fields can be decomposed
into left and right moving parts:
\barray
\label{3.3}
\bfphi^a (z ,\zbar ) &=& \phi^a (z) + \phibar^a (\zbar) 
\\ \nonumber
\bfchi^a (z, \zbar) &=& \chi^a (z) + \chibar^a (\zbar) 
\earray
In the sequel, we will continue to display local fields in {\bf bold}
face.   The two point functions are
\beq
\label{3.4}
\langle \phi^a (z) \phi^b (w) \rangle = - \eta^{ab} \log (z-w), 
~~~~~
\langle \chi^a (z) \chi^b (w) \rangle = - \ep^{ab} \log(z-w) 
\eeq
(Our conventions are $\eta^{ab} = \eta_{ab}, \ep^{ab} = \ep_{ab}$.)

It is straightforward to verify the following representation of
the OPE's in eq. (\ref{2.4}):
\barray
\label{3.5}
H &=& i \sqrtk \, \d_z \phi^1 , ~~~~~ J = i \sqrtk \, \d_z \phi^2 
\\ \nonumber
S_+ &=& \sqrtk \, \d_z \chi^1\,  e^{i
 (\phi^1 - \phi^2 )/\sqrtk }
,
~~~~~
S_- = - \sqrtk \, \d_z \chi^2 \, e^{ -i 
 (\phi^1 - \phi^2 )/\sqrtk  }
\earray
In the sequel, where there is no cause for confusion, we
will simply write $ \d \phi$ for $\d_z \phi (z)$.

\section{Finite dimensional representations of $gl(1|1)$.}

The complete solution of the current algebra as a quantum
field theory requires the determination of the spectrum
of fields.  The chiral primary fields $V_\rep (z)$ transform
as finite dimensional representations $\rep$  of $gl(1|1)$, which
is equivalent to the OPE:
\beq
\label{4.1}
J^a (z) \, V_\rep (0) \sim  \inv{z} ~  t^a_\rep \, V_\rep (0) 
\eeq
where $J^a, a=1,.,4$ are the $\glk$ currents and $t^a_\rep$ is 
the  finite dimensional matrix representation of $r$ of  $gl(1|1)$.   
(In the sequel we will continue to refer to general super-currents
as $J^a$.)   

Before explicitly constructing  the primary fields
$V_\rep$, we first describe the relevant finite dimensional
representations\cite{Kac,Gotz2,Schomerus1}. 
  The $gl(1|1)$ algebra has  the following
non-zero (anti) commutation relations:
\beq
\label{4.2}
[H, S_\pm ] = [J, S_\pm ] = \pm S_\pm , ~~~~~
\{ S_+, S_- \} = H-J 
\eeq
The fermionic operators are nilpotent: $S_\pm^2 = 0$.  
(It is implicit that the above generators are the zero modes
of the currents.) 
First, there are one-dimensional representations where
$S_\pm = 0$,  $H=J=h$.  We will denote these as $\hone h$.  

The so-called typical representations are two-dimensional:
\barray
\label{4.3}
H &=& \(\matrix{h&0\cr 0& h-1\cr }\) , ~~~~~
J= \(\matrix{j&0\cr0&j-1\cr}\) 
\\ 
S_+ &=& \( \matrix{ 0&b\cr0&0\cr } \), ~~~~~
S_- = \( \matrix{ 0&0\cr c&0\cr } \) 
\earray
where $bc=h-j$.   Let us denote these representations
as $\glrep{h}{j}$.  When $h\neq j$ these representations 
are irreducible.  
The tensor product of two typical representations can be 
deduced by simply considering the $U(1)$'s:
\beq
\label{4.4}
\glrep{h_1}{j_1} \otimes \glrep{h_2}{j_2} = 
\glrep{h_1 + h_2}{j_1 + j_2 } \oplus
\glrep{h_1 + h_2 -1}{j_1 + j_2 -1} 
\eeq

When $h=j$,  the representations are reducible but
indecomposible.   There are two different representations
depending on whether $b$ or $c$ equals zero.  
For $b=0$, the representation will be referred to as 
$\glrep{h}h$ and for $c=0$ as $\glrepprime{h}$.   They can
be reduced as $\glrep{h}h = \hone{h} \oplus  \hone{h-1}$
however  they are indecomposable since $S_- : ~~\hone{h} \to 
\hone{h-1}$.

Finally there are 4-dimensional indecomposable representations
which will be important in the sequel, and 
we  denote as $\hfour{h}$.   
They arise in the tensor product of typical representations
when $h_1 + h_2 = j_1 + j_2$:
\beq
\label{4.5}
\glrep{h_1}{j_1} \otimes \glrep{h_2}{j_2} = 
\hfour{h_1 + h_2 -1}
\eeq
From eq. (\ref{4.4}), one sees that $\hfour{h}$ can be reduced
into
$\glrep{h+1}{h+1} \oplus \glrep{h}{h}$, however $S_\pm$ 
mixes these two representations.   The generators
in $\hfour{h}$ are:
\beq
\label{4.6}
S_+  = \( \matrix{
0 & 0 & 1 & 0 \cr
0 & 0 & 0 & -1\cr
0 & 0 &0  & 0 \cr
0 & 0 &0  & 0 \cr
} \) , ~~~~~
S_- = \( \matrix{
0 & 0 &0 & 0 \cr
-1 & 0 &0  & 0 \cr
0 & 0 &0  & 0 \cr
0 & 0 &-1  & 0 \cr
} \)
, ~~~~~
H  = J = \( \matrix{
h+1 & 0 &0  & 0 \cr
0 & h &0  & 0 \cr
0 & 0 &h  & 0 \cr
0 & 0 &0  & h-1 \cr
} \) 
\eeq
The indecomposability  can be represented by the diagram in
Figure 1.

\begin{figure}[htb] 
\begin{center}
\hspace{-15mm}
\includegraphics[width=10cm]{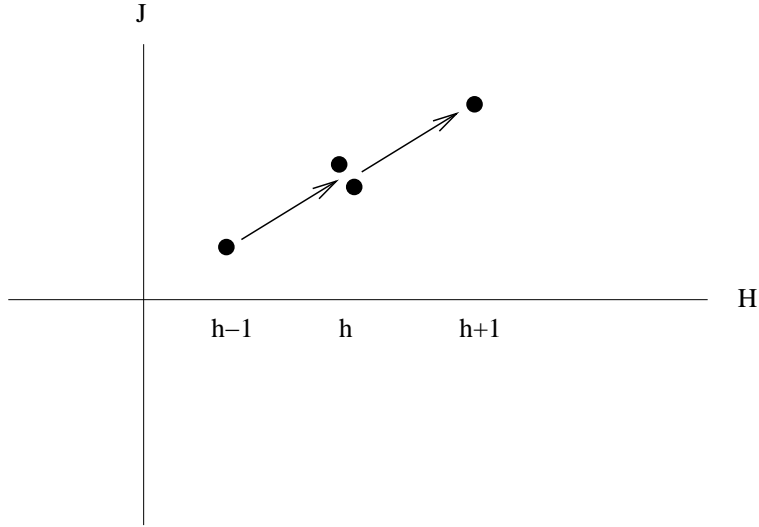} 
\end{center}
\caption{The four dimensional indecomposable representation 
$\hfour{h}$ of $gl(1|1)$.  The arrows indicate the action of
$S_+$.} 
\vspace{-2mm}
\label{Figure1} 
\end{figure}

If there exists primary fields corresponding to the 
representation $\rep$,  then the conformal scaling
dimension $\Delta_\rep$ follows from the
Sugawara form eq. (\ref{glstress}) and the values of the
casimirs $C_2 , C_2'$ in the representation $\rep$.  This
way one finds
\beq
\label{4.9}
\Deltahj{h}{j} = \frac{ (h-j)^2 }{2k^2} + \frac{ (h-j) (h+j -1)}{2k} 
\eeq
and $\Delta_{\hone{h}} = \Delta{_\glrep{h}{h}}  = \Delta_{\glrepprime{h}} = 0$. 
Note that $\Deltahj{h}{j}(k) = \Deltahj{j}{h} (-k)$,  in 
accordance with the automorphism eq. (\ref{auto}). 

The novel feature of the representation $\hfour{h}$ is that 
the casimir $C_2$ is not diagonal:
\beq
\label{4.10}
C_2 = \( \matrix{
0 & 0 &0  & 0 \cr
0 & 0 &2  & 0 \cr
0 & 0 &0  & 0 \cr
0 & 0 &0  & 0 \cr
} \) 
\eeq
As we will see, this leads to logarithmic properties
of the fields,  as explained in \cite{Serban} for $osp(2|2)$.

The original fields in the $k=1$ representation of section II
correspond to:
\beq
\label{original}
(\psi_+ , \beta_+ ) \leftrightarrow \glrep10, ~~~~~
(\beta_- , \psi_- ) \leftrightarrow \glrep01
\eeq
A consistency check is $\Deltahj{1}0 = \Deltahj{0}1 = \inv{2}$ 
when $k=1$.

\section{Twist and logarithmic operators in the symplectic fermion theory.}

In this section we present the two important features of symplectic fermions
we will need in order to construct the primary fields of the current algebra,  
namely the logarithmic and twist fields.     
More details are provided in Appendix B.

\subsection{Logarithmic fields} 

The $c=-2$ symplectic fermion theory is the simplest and most studied
example of a logarithmic conformal field theory\cite{Gurarie,Kausch}.  
For a review see\cite{Flohr,Gaberdiel}.   The original investigation of these
properties was based on the null vector differential equation for the
$c=-2$ minimal model 4-point functions\cite{Gurarie}.   In this abstract
approach,  the explicit construction of the logarithmic operators is not evident.
The logarithmic operator is also not contained in the first order
$\eta-\xi$  description\cite{Kausch}.  (See Appendix B.)    An important
feature of the second order $\chi$ description is that the logarithmic fields
are explicitly contained in the theory because of the additional zero modes.

Generally, 
let $\ell_0 (z),  \ell (z)$ denote a logarithmic pair with scaling dimension $\Delta$.  
By definition they satisfy the following OPE with the stress tensor:
\barray
\label{5.1}
T(z) \, \ell_0 (0) & \sim &  \frac{\Delta}{z^2} ~ \ell_0 (0) + \inv{z}  \, \d \ell_0 (0) 
\\ \nonumber
T(z) \, \ell (0)    &\sim & \frac{\Delta}{z^2} \, \ell (0) + \frac{a}{z^2} \, \ell_0 (0)    + \inv{z} \d \ell (0) 
\earray
which implies
\beq
\label{5.2}
L_0 | \ell_0 \rangle = \Delta | \ell_0 \rangle , ~~~~~ 
L_0 |\ell \rangle =  \Delta | \ell\rangle + a | \ell_0 \rangle 
\eeq
The conformal Ward identities lead to the following two point 
functions\cite{Caux1,Serban}:
\beq
\label{5.3}
\langle \ell_0 (z) \ell_0 (0) \rangle = 0, ~~~~~ 
\langle \ell (z) \ell_0 (0) \rangle = \frac{C}{z^{2 \Delta}}  , ~~~~
\langle \ell (z) \ell (0) \rangle = \frac{C' - 2a C \log z}{z^{2\Delta}} 
\eeq
where $C, C'$ are constants. 

The stress tensor of the symplectic fermion is
\beq
\label{5.4}
T(z) = \inv{2} \ep_{ab}  \d \chi^a \d \chi^b  
\eeq
Define
\beq
\label{5.5}
\ell (z) = -\inv{2} \ep_{ab}  \chi^a \chi^b =  - \chi^1 \chi^2 
\eeq
Then $\ell (z)$ and $\ell_0 = 1$ satisfy the OPE's eq. (\ref{5.1})  with $\Delta = 0$.
On the states $|\ell\rangle = \ell (0) |0\rangle$
 and $|0\rangle = |\ell_0 \rangle$,  
the Virasoro zero mode $L_0$ has the usual form for
a logarithmic pair:   $L_0 = \( \matrix{ 0 & 1\cr 0&0&\cr } \) $.  

The field can be expanded as follows:
\beq
\label{5.6}
\chi^a (z) = \chi^a_0 - i \chitilde^a_0 \, \log (z) + i 
\sum_{n\neq 0} \inv{n} \, \chi_n^a \, z^{-n} 
\eeq
where $\{ \chi_0^a , \chitilde_0^b\}  =i \ep^{ab} $ and
$\{ \chi_m^a, \chi_n^b \} = m  \ep^{ab} \delta_{m+n, 0} $.    
In terms of the zero modes:
\beq
\label{5.7b}
| \ell \rangle = - \chi^1_0  \chi^2_0  |0\rangle 
\eeq
Functional integrals over $\bfchi$ are zero
unless they contain  the zero modes:
\beq
\label{5.7c}
\langle \ell \rangle = 1 , ~~~~~ \langle 1 \rangle = 0
\eeq
consistent with eq. (\ref{5.3}).

\subsection{Twist fields.}

As for the spin fields of the Ising model,  the twist fields modify the boundary
conditions of the fundamental field $\chi$:
\barray
\label{5.8}
\chi^1 ( e^{2\pi i } z )  \mu_\lambda (0) &=&e^{-2\pi i \lambda} \chi^1 (z) \mu_\lambda (0) 
\\ \nonumber
\chi^2 ( e^{2\pi i } z )  \mu_\lambda (0) &=&e^{2\pi i \lambda} \chi^2 (z) \mu_\lambda (0) 
\earray
The properties of these fields were studied in\cite{Kausch}.  
It is clear from the above equation that $2 \pi \lambda$ is a phase and is restricted to
$-1 < \lambda < 1 $.    In the presence of $\mu_\lambda$,  the mode expansion 
is twisted:
\barray
\label{5.9}
\chi^1 (z) &=& \chi^1_0 + i  \sum_{n \in \Zmath}   \inv{n+\lambda}   \chi^1_{n+\lambda}  ~ z^{-n-\lambda} 
\\ \nonumber 
\chi^2 (z) &=& \chi^2_0 + i  \sum_{n \in \Zmath}   \inv{n-\lambda}   \chi^2_{n-\lambda}  ~ z^{-n+\lambda} 
\earray
The expansion eq. (\ref{5.6}) arises as $\lambda \to 0$.   

We will need the OPE of the twist fields with $\d \chi$,  which involves new fields 
$\sigma_\lambda^a$.   The results are derived in Appendix B:
\barray
\label{5.10}
\d \chi^1 (z) \,  \mu_\lambda (0)  &\sim & \frac{ \sqrt{1-\lambda}}{z^\lambda} ~ \sigma^1_\lambda (0),  
~~~~~
\d \chi^2 (z)  \, \sigma_\lambda^1  (0) \sim \frac{ \sqrt{1-\lambda}}{z^{2-\lambda}} ~ \mu_\lambda (0) 
\\ \nonumber
\d \chi^2 (z) \,  \mu_\lambda (0)  &\sim&  \frac{  \sqrt{\lambda}} {z^{1-\lambda} }  ~ \sigma_\lambda^2 (0) , ~~~~~
\d \chi^1 (z) \, \sigma_\lambda^2 (0)  \sim  -  \frac{\sqrt{\lambda}}{z^{1+\lambda}}  ~ \mu_\lambda (0) 
\earray
The scaling dimensions of these fields is
\beq
\label{5.11}
\Delta (\mu_\lambda)  = \Deltachi_\lambda , ~~~~~ \Delta(\sigma_\lambda^1 ) = \Deltachi_{\lambda -1} , 
~~~~~ \Delta( \sigma_\lambda^2 ) = \Deltachi_{\lambda +1} 
\eeq
where we have defined:
\beq
\label{5.12} 
\Deltachi_\lambda \equiv   \frac{\lambda (\lambda -1)}{2}  
\eeq
The powers of $z$ in eq. (\ref{5.10}) are fixed by these scaling dimensions.   

The factors of $\sqrt{\lambda}, \sqrt{1-\lambda}$  are not arbitrary and will be important in
the next section.   They are fixed once the normalizations 
$\langle \mu_{1-\lambda} | \mu_\lambda \rangle =1$ and
$\langle \sigma_{1-\lambda}^a | \sigma^b_\lambda \rangle = \ep^{ab}$ are fixed.
(See Appendix B.)

\section{Vertex operators.}

In this section we explicitly construct the chiral (left-moving) vertex operators for
the $gl(1|1)$ representations in section IV.   We present formulas for $k>0$;  negative
$k$ results follow from the $k\to -k$ automorphism (\ref{auto}).   For a general
current $J^a$, the vertex operators for a finite dimensional representation $\rep$ are
a vector of fields $V_\rep^\alpha (z)$,  
$\alpha = 1, 2, .., {\rm dim} (\rep)$ satisfying the 
OPE
\beq
\label{6.1} 
J^a (z) \,  V^i_\rep (0) = \inv{z} ~ t^a_{j i} V^j_\rep
\eeq
where $t^a$ is the finite dimensional matrix representation of $\rep$.  

Introduce the notation for the bosonic sector:
\beq
\label{6.2}
\Vbos{h}{j} \equiv  e^{i (h\phi^1 - j \phi^2 )/\sqrtk } 
\eeq
The above field has $U(1)$ charges $(H,J) = (h,j)$ 
and conformal dimension 
\beq
\label{6.3}
\Delta_{h,j}^\phi = \frac{h^2 - j^2}{2k} 
\eeq
The vertex operators $V_{\hone{h}}$ for the one-dimensional representation
$\hone{h}$ are purely bosonic:
\beq
\label{6.4}
V_{\hone{h}} = \Vbos{h}{h}
\eeq

Let $\V {h}j$ denote the vertex operator for the 2-dimensional typical representation 
with $h\neq j$.   They require the twist fields with $\lambda = \frac{h-j}{k}$.
For $h>j$ one has:
\beq
\label{6.5}
\V{h}j  = (h-j)^{1/4} \( \matrix{
-   \mu_\lambda ~  \Vbos{h}{j}  \cr
\sigma^2_\lambda ~ \Vbos{h-1}{j-1} \cr }
\) , ~~~~~ \lambda = \frac{h-j}{k}
\eeq
For $h<j$ the proper expression is
\beq
\label{6.6}
\V{h}{j} = (j-h)^{1/4} \( \matrix{
\sigma_{1+\lambda}^1 ~ \Vbos{h}{j}  \cr
\mu_{\lambda+1} ~ \Vbos{h-1}{j-1} \cr }  \) 
, ~~~~~ \lambda = \frac{h-j}{k}
\eeq
To verify that these expressions satisfy eq. (\ref{6.1}), 
one uses the explicit expressions for
the $\glk$ currents (\ref{3.5}),  
the representations $t^a$ given in section III,  and the OPE's 
(\ref{5.10}).    In doing so,  one finds that the factors of $\sqrt{\lambda}, \sqrt{1-\lambda}$ 
in the OPE's (\ref{5.10}) are necessary.    
The construction is also consistent with the scaling dimension $\Deltahj{h}{j}$ in
eq. (\ref{4.9}):
\beq
\label{6.7}
\Deltahj{h}{j}= \Delta^\phi_{h,j}  + \Deltachi_\lambda = \Delta^\phi_{h-1, j-1} + \Deltachi_{\lambda+1}
, ~~~~~\lambda = \frac{h-j}{k}
\eeq
When $h=j$, the vertex operators for the representations $\glrep{h}{h}$ and
$\glrepprime{h}$ are 
\beq
\label{6.8}
\V{h}{h}  = \( \matrix{   \chi^1 ~ \Vbos{h}{h }\cr
-\sqrtk ~ \Vbos{h-1}{h-1}   \cr  } \), ~~~~~
V_{\glrepprime{h}} = \(  \matrix{ 
- \sqrtk ~ \Vbos{h}{h} \cr 
\chi^2  ~ \Vbos{h-1}{h-1}  \cr  } \) 
\eeq

The vertex operators for $\hfour{h}$ are novel because they are logarithmic.   
The zero modes of the $\chi$ fields span a 4-dimensional vector space
$|0\rangle, \chi^1 |0\rangle , \chi^2 |0\rangle,  \chi^1 \chi^2 |0\rangle$, 
and the vertex operator is built on this structure:  
\beq
\label{6.9}
\Vfourh = \( \matrix{ 
\chi^1 ~ \Vbos{h+1}{h+1}  \cr
\sqrtk ~ \Vbos{h}h \cr
 \chi^1 \chi^2 ~ \Vbos{h}{h} /\sqrtk\cr
\chi^2 ~ \Vbos{h-1}{h-1}  \cr } \)
\eeq
The two middle fields $\ell'_0 = \sqrtk \Vbos{h}h$, 
$\ell' = \chi^1 \chi^2 \Vbos{h}{h} /\sqrtk $ form a logarithmic
pair (\ref{5.1}) with $\Delta=0$ and $a=-1/k$  since $\ell_0 =1$
and $\ell (z) $ in eq. (\ref{5.5}) form such a pair.  
As explained in \cite{Serban}, this logarithmic property is
reflected in the fact that the casimir $C_2$ is not diagonal for 
$\hfour{h}$,  eq. (\ref{4.10}).   Using the Sugawara form 
(\ref{glstress}) and eq. (\ref{4.10}), one indeed sees that on
$\hfour{h}$:
\beq
\label{6.11}
L_0 = - \inv{k} \( \matrix{
0&0&0&0\cr
0&0&1&0\cr
0&0&0&0\cr
0&0&0&0\cr
} \)
\eeq
where $L_0$ is the zero mode of $T(z)= \sum_n L_n z^{-n-2} $, 
which is consistent with the form of the vertex operator 
 eq. (\ref{6.9}).

\section{Closed operator algebras and the spectrum of fields}

As for ordinary current algebras,  not all representations of
$gl(1|1)$ correspond to primary fields.  For example, for
$su(2)_k$,  only the primary fields with spin $j\leq k/2$
are present in the spectrum\cite{Gepner}.    For $\glk$ there are similar
restrictions depending on the level $k$.  Since the twist
fields $\mu_\lambda$ are defined for $-1 \leq \lambda \leq 1$ 
and the vertex operators $\V{h}{j}$ involve $\lambda = \frac{h-j}{k}$,
it is clear that:
\beq
\label{7.1}
- k \leq h-j \leq k 
\eeq

The above restriction can also be understood directly in the
affine super-algebra.  Let $|h,j\rangle_{hw}$ denote
a highest weight state satisfying
\beq
\label{7.2}
S^\pm_n  |h,j\rangle_{hw} = S_0^+ | h,j\rangle_{hw} = 0, ~~~~n >  0 
\eeq
Consider the modes $S^\pm_{1} , S^\pm_{-1}$ which satisfy
two $gl(1|1)$'s:
\beq
\{ S^+_1 , S^-_{-1} \} = k + H_0 - J_0 , ~~~~~
\{ S^+_{-1} , S^-_1 \} = -k + H_0 - J_0 
\eeq
Then one has
\beq
\label{7.3}
\{ S^+_{-1} , S^-_1 \} | h,j\rangle_{hw} =
S_1^- S^+_{-1} | h,j\rangle_{hw} =  (h-j-k) |h,j\rangle_{hw} = 0
\eeq
This means there is a  null state
 $S^+_{-1} | h,j\rangle_{hw} = 0$ if $h-j = k$.  
Using this null state inside a 3-point function one
deduces that the primary fields must satisfy eq. (\ref{7.1}).  
The fusion rules are as in eq. (\ref{4.4}) where only
fields satisfying $h-j \leq k$ are kept on the right hand side.

Thus far there is no restriction on the level $k$.  The manner
in which $k$ enters the vertex operator construction shows that in
the bosonic sector $\Vbos{h}{j}$, $k$ can be 
interpreted as a radius of compactification $R=\sqrtk$.   

For generic irrational $k$, one does not have a closed operator
algebra.    A closed operator algebra is obtained when $k$ is
an integer and $(h,j)$ are integers.   These are the ``minimal
models''  based on $\glk$.  This situation arises
naturally in the application to disordered systems since the
fundamental fields $\psi_\pm , \beta_\pm$  at $k=1$ correspond to the doublets 
$\glrep{1}0$ and $\glrep{0}1$, eq. (\ref{original}).  
It will be shown in the next section how one can obtain
higher integer $k$ in the multi-copy theory via a
super spin-charge separation.   Thus it appears the
locality considerations in \cite{Mudry1}, which led to
the restriction $k=1/m$ with  $m$ is an integer, is too restrictive.  

The closed operator algebra at higher integer level $k$ is
generated by repeated OPE of the two vertex operators
$\V{1}{0}$ and $\V{0}1$, and are subject to the restriction 
eq. (\ref{7.1}).  Note that the scaling dimensions
follow the pattern
\beq
\label{7.5}
\Deltahj{h+n}{j+n} = \Deltahj{h}j + \frac{n(h-j)}{k} 
\eeq

\subsection{The case of $k=2$.}

Let us illustrate these features in the next simplest case of $k=2$.  
The twist fields $\muhalf$ and $\sigmahalf^a$ have 
$\Delta$ equal to $-1/8$ and $3/8$ respectively.   These
fields have the following OPE\cite{Kausch}
\barray
\nonumber
\muhalf (z) \, \muhalf (0) &=& z^{1/4} \( \ell (0) + \log (z) + ....\) 
\\ \label{7.6}
\sigmahalf^a (z) \, \sigmahalf^b (0) &=& \inv{z^{3/4}} \ep^{ab} 
\( \ell (0) + \log (z) + ... \) 
\\ \nonumber
\muhalf(z) \, \sigmahalf^a (0) &=& -\inv{\sqrt{2}} \inv{z^{1/4} } 
\( \chi^a (0) + ... \) 
\earray
The vertex operators $\V{1}0$ and $\V{0}1$ both have conformal
dimension $1/8$ and take the form
\beq
\label{7.7}
\V{1}0 = \( \matrix{ - \muhalf \, \Vbos{1}0  \cr 
\sigmahalf^2 \, \Vbos{0}{-1} \cr } \) 
,~~~~~
\V{0}1 = \( \matrix{ \sigmahalf^1 \, \Vbos{0}{1} \cr
\muhalf \, \Vbos{-1}{0} \cr } \)
\eeq
Using the OPE's (\ref{7.6}) one finds
\beq
\label{7.8}
\V{1}{0} (z) \, \V{0}{1} (0) \sim \inv{z^{1/4}} 
\Vfour{0} 
\eeq
where $\Vfour{0}$ is the 4-dimensional logarithmic field
eq. (\ref{6.9}).   

To find the other OPE's, we need the $\lambda = 0,1$ limit of
the twist fields in the expressions (\ref{6.5},\ref{6.6}) for
the vertex operators $\V{2}{0}, \V{1}{-1}, \V{0}{2}, \V{-1}{1}$. 
The following linear combinations 
are consistent with the $\lambda = 0,1$ limit of the OPE's
in eq. (\ref{5.10}):
\barray
\label{7.8b}
\mu_1 &=& a + b \, \chi^2 , ~~~~~ 
\sigma_1^2 = a\,  \d \chi^2 + b\,  \d \chi^2 \, \chi^2 
\\ \nonumber 
\mu_0 &=& c + d\,  \chi^1 , ~~~~~ \sigma_0^1 = c \, \d  \chi^1 + d\,  \d\chi^1 \, \chi^1
\earray
where $a,b,c,d$ are constants.   Which linear combinations 
appear in the vertex operators follows from eq. (\ref{7.6}).  
One finds
\barray
\label{7.9}
\V{1}{0} (z) \, \V{1}{0} (0) &\sim& 
 z^{-1/4} ~ \V{1}{-1}   +  z^{3/4} ~ \V{2}{0} 
\\ \nonumber
\V{0}{1} (z) \, \V{0}{1} (0) &\sim& 
z^{-1/4} ~ \V{0}{2} + z^{3/4} ~ \V{-1}{1}  
\earray
where 
\barray
\nonumber
\V{2}{0} &=& \sqrt{2} \( 
\matrix{ - \Vbos{2}{0} \cr \d\chi^2 \Vbos{1}{-1} \cr }  \) , 
~~~~~ \V{1}{-1} = \sqrt{2} \( 
\matrix{ - \chi^2 \Vbos{1}{-1} \cr 
\d \chi^2 \chi^2 \Vbos{0}{-2} \cr } \) 
\\&~& \label{7.10}  \\ \nonumber
\V{0}{2} &=& \sqrt{2} \( \matrix{
\d\chi^1  \chi^1  \Vbos{0}{2} \cr
\chi^1 \Vbos{-1}{1}  \cr } \) , ~~~~~
\V{-1}{1} = \sqrt{2} \( \matrix{
\d\chi^1 \Vbos{-1}{1} \cr \Vbos{-2}{0} \cr } \) 
\earray

The remaining low dimension fields are $\V{2}{1}$ 
and $\V{1}{2}$ with $\Delta = 5/8, -3/8$ respectively.  
By virtue of eq. (\ref{7.5}), the other fields have
dimension which differs by an integer from the fields
considered thus far.

Since $\glhat_2$ has $c=0$,  it is interesting to compare it with
the $c=0$ minimal Virasoro model.  Let us refer to the minimal
model fields at $c=0$ as $\Phi_{m,n}$ with conformal dimension
$\Deltamin_{m,n}$.  (See \cite{CFT,ginsparg}.)  The two models share the dimensions
$1/8$ and $5/8$ since $\Deltamin_{2,2} = 1/8$ and
$\Deltamin_{2,1} = 5/8$.    The latter determines the 
correlation length exponent for percolation, 
$\nu_{\rm perc.} = (2(1-\frac{5}{8}))^{-1} = 4/3$.   Note that the 
field $\Phi_{1,3}$ with $\Delta = 1/3$ is not present
in the $\glhat_2$ theory.   The field $\Phi_{1,3}$ is known
to determine the correlation length exponent for 
self-avoiding walks, $\nu_{SAW} = 3/4$.  
Thus, as a possible disordered critical point, $\glhat_2$ 
is more closely related to percolation.    However it
is not entirely equivalent to it since it does not
contain for example all the hull exponents considered 
in\cite{Duplantier}.    We will
return to this point in section XII  where  we discuss
applications to the spin quantum Hall  transition.

\section{Super spin-charge separation and disordered critical points.}

Spin-charge separation for ordinary $su(2)$ Dirac fermions
has many important applications, for example to Luttinger
liquids in $1d$.   In this section we present the extension
of this construction to super current algebras.   
We also  explain  how the higher level $\glk$ theory can arise
as a disordered critical point.  

Consider the action (\ref{2.1}) for Dirac fermions only,  extended
to $N$-copies:
\beq
S^{\rm N-copy} = \inv{4\pi} \int d^2 x  
\sum_{\alpha=1}^N \(  \psi^\alpha_-  \d_\zbar \psi_+^\alpha + 
\psibar_-^\alpha \d_z \psibar_+^\alpha \) 
\eeq
The model now has an $\suNhat{k=1}$ symmetry with currents 
\beq
\label{8.1}
L^a_\psi = \psi_-^\alpha t^a_{\alpha\beta} \psi_+^\beta
\eeq
where here $t^a$ are a matrix representation of the vector 
of $su(N)$.  The model also has a $u(1)$ symmetry which commutes
with $su(N)$.    Spin-charge separation is the statement
that the full stress tensor for the free theory can be decomposed
into commuting parts: 
\beq
\label{8.2}
T^\Ncopy_\free = - \inv{2} \sum_\alpha \psi_-^\alpha \d_z \psi_+^\alpha 
= T_{u(1)} + T_{\suNhat{1}} 
\eeq
where $T_{\suNhat{1}}$ is the Sugawara stress tensor and $T_{u(1)}$ 
is the stress tensor for a single scalar field. (See for 
instance \cite{ginsparg}.)  A check of
the above decomposition is the central charge. 
The $\suNhat{k}$ theory has $c_{\suNhat{k}} = \frac{k(N^2-1)}{(k+N)}$,
whereas the $u(1)$ has $c=1$.   When $k=1$, the total $c$ equals $N$, 
as appropriate
for $N$ Dirac fermions.   The other check involves the scaling
dimension.  The $N$-dimensional vector  representation at level $k$ has
\beq
\label{vecsun}
\Delta_{\suNhat{k}} = \frac{N^2-1}{2N(k+N)}
\eeq
The $u(1)$ is at 
radius $R=\sqrt{N}$,  with $\Delta_{u(1)} = \inv{2N}$, and one
verifies
$\Delta(\psi_\pm) = \Delta_{u(1)} + \Delta_{\suNhat{1}} = \inv{2}$.

Consider now $N$ copies  of the theory (\ref{2.1}) with ghosts
$\beta_\pm^\alpha$.  This theory has the maximal $osp(2N|2N)_1$
symmetry.    In the ghost sector the currents
\beq
\label{8.4}
L^a_\beta = \beta_- t^a \beta_+ 
\eeq
satisfy $\suNhat{k=-1}$.   We will need the following basic
result.   Given two  copies of the same current 
algebra with currents $J^a_1$ at level $k_1$ and $J^a_2$
at level $k_2$ which furthermore commute, $[J^a_1 (z) , 
J^b_2 (w) ] = 0$.   Then $J^a = J^a_1 + J^a_2$ 
satisfies the current algebra at level $k_1 + k_2$.  
The complete $su(N)$ currents
$L^a = L^a_\psi + L^a_\beta$ thus have level $k=0$.

The model also has a
$\glhat$ symmetry generated by the currents:
\beq
\label{8.5}
H = \sum_\alpha \psi_+^\alpha \psi_-^\alpha , ~~~~
J = \sum_\alpha \beta_+^\alpha \beta_-^\alpha , ~~~~
S_\pm = \pm \sum_\alpha \psi_\pm^\alpha \beta_\mp^\alpha
\eeq
Since the above currents are sums of the currents in each copy 
with level $k=1$,
they satisfy $\glk$ with $k=N$. 
It is also important that these  $\glhat_N$ currents
commute with the $\suNhat{k=0}$.  
The super spin-charge separation is the non-trivial statement:
\beq
\label{8.6}
T^\Ncopy_\free = 
 - \inv{2} \sum_\alpha (\psi_-^\alpha \d_z \psi_+^\alpha 
 + \beta_-^\alpha \d_z \beta_+^\alpha ) =  
T_{\glhat_{N}} + T_{\suNhat{0}} 
\eeq
As we will show in section X,  $T_{\glhat_2} = T_{\osphat_{-2}}$,
and this form of the relation (\ref{8.6}) was proved 
for $k=2$ in \cite{SpinCharge}.  See also \cite{Bhaseen}.
The more general relation above for any $N$ can be proved similarly.  
Note that since both current algebras have $c=0$,  this is
consistent with $c_\free = 0$.   A more non-trivial
check at arbitrary $N$ is based on the conformal  dimensions.  The fields
$(\psi_+ , \beta_+ ), (\beta_- , \psi_-)$ transform in the
$\glrep{1}{0}, \glrep{0}{1}$ representations of $\glhat_N$ with
$\Deltahj{1}{0} =\Deltahj{0}{1}= \inv{2N^2}$.  
The vector  representation of
$\suNhat{0}$ has $\Delta_{\suNhat{0}} = \frac{N^2 -1}{2N^2}$ so that
\beq
\label{8.7}
\Delta(\psi_\pm, \beta_\pm ) = \Delta_{\glrep{1}{0}}  
+ \Delta_{\suNhat{0}} = \inv{2}
\eeq

The $\glhat_N$ theory can arise as a disordered critical point
as follows. 
More generally,  consider two commuting current algebras
$\CG_A$ and $\CG_B$ with currents $J_{A},  J_{B}$.
Furthermore, let us suppose that the stress tensor for a given
conformal theory separates  as in eq. (\ref{8.6}).  
Consider the perturbation of the conformal field theory
by left-right current-current perturbations:
\beq
\label{curr.1}
S = S_{\rm cft}  +  \int \frac{d^2 x}{2\pi}  \(  
g_A  \, J_{A} \cdot \bar{J}_{A} +  
g_B  \, J_{B} \cdot \bar{J}_{B}
\)
\eeq
where $J\cdot \bar{J}$ is the invariant built on the
quadratic casimir.   Since the currents commute,  the
renormalization group (RG)  beta-functions decouple;  to 1-loop the result is: 
\beq
\frac{d g_A }{d \ell} = C^{\rm adj}_A \, g_A^2 , ~~~~~
\frac{d g_B}{d \ell} = C^{\rm adj}_B \, g_B^2
\eeq
where $\ell$ is the logarithm of the length scale
and $C^{\rm adj}_A $ is the casimir for the adjoint representation 
of $\CG_A$.   Let us suppose that the physical
regime corresponds to positive $g_{A,B}$.  
If  $C^{\rm adj}_B$ is positive, then the  coupling
$g_B$ is marginally relevant and the flow is to infinity. 
This is a massive sector as in the Gross-Neveu model. 
These massive  $\CG_B$ degrees of freedom
are decoupled at low energies.  We will refer to 
the $\CG_B$ degrees of freedom as being ``gapped-out'' in
the RG flow to low energies.   
If $C^{\rm adj}_A$ is negative,  then the  coupling $g_A$ is 
 marginally irrelevant.  This results in the fixed point
defined by the theory with current algebra symmetry 
$\CG_A$.  If the original conformal field theory 
corresponds to the current algebra $\CG_{\rm max}$,   then the 
fixed point  may be viewed as  the coset
$\CG_{\rm max}/\CG_B$.  For $N$-copies of Dirac fermions
and ghosts,  $\CG_{\rm max} = osp(2N|2N)_1$.  This scenario was proposed for 
generic fixed points of marginal current-current perturbations in
\cite{LeClairSC},  however here it is a somewhat trivial example of
the GKO construction\cite{coset} because of the decomposition 
of the stress tensor.    In fact, what was missing in
the arguments in \cite{LeClairSC} was precisely  the spin-charge
separation.

Returning to disordered Dirac fermions,  
the N-copy version is relevant for the computation 
of averages of multiple moments (multi-fractality) or can be
part of the definition of the 1-copy theory, as in the spin
quantum Hall transition which has an $su(2)$ symmetry from the very beginning
and thus the  $1$-copy theory  corresponds to $N=2$. 
Disorder averaging generally leads to left-right current-current
perturbations.    For certain kinds of disorder,
where perhaps some of the disorder is set to zero, 
the disorder averaged effective action takes the form
(\ref{curr.1}) with $\CG_A = gl(1|1)_N$ and $\CG_B = su(N)_0$. 
The  $su(N)_0$ current interactions can arise from 
a disordered $su(N)$ gauge field, but not necessarily so;
other scenarios will be described in\cite{SuperDisorder}.  

For the $su(N)_0$ currents $L^a$, $C^{\rm adj} >0$.  For
super-current algebras like $osp(2N|2N)$, $C^{\rm adj} < 0$. 
For $gl(1|1)$ the situation is somewhat more subtle because
there are two quadratic casimirs\cite{Guruswamy}.  Consider
\beq
\label{fac2}
S = S_{gl(1|1)_k} + \int \frac{d^2 x}{2\pi}  \( 
g  \, \( J\bar{J} - H\bar{H} + S_+ \bar{S}_- - S_- \bar{S}_+ \) 
+ g'  (J-H)(\bar{J} - \bar{H} ) \)
\eeq
where $S_{gl(1|1)_k}$ formally represents the conformal theory
with $gl(1|1)_k$ symmetry.  The latter can be taken to have the
 free field form eq. (\ref{3.1}).  
Then the 1-loop beta function for $g$ is zero, 
whereas $d g'  / d \ell = - g^2 $.    Therefore the
$gl(1|1)$ current interactions are marginally  irrelevant.   
This is to be contrasted with the situation for the 
model in \cite{Guruswamy} since there $g'$ corresponded
to the variance of disordered {\it imaginary} $u(1)$ gauge field
and this changes the sign of the coupling.  
The higher loop corrections computed in \cite{LeClairSC} 
do not alter this picture.  

For $N=2$ the analog of this  flow to $\osphat_{-2}$ for
pure  $su(2)$ gauge  disorder 
was proposed in \cite{SpinCharge}.
In \cite{Tsvelik2}  the  $\osphat_{-2}$ description of 
strongly disordered gauge fields was 
shown to be consistent with other approaches such
as \cite{Bernard1,Mudry1}.  
More interesting models, such as the spin quantum Hall transition, 
have additional kinds of disorder besides pure gauge field disorder
and thus should correspond to relevant perturbations of the 
current algebra. 
We will return to this issue  in section XII, where we show that
the perturbation is by a logarithmic operator.

\section{Local $gl(1|1)$ invariant operators}

The left and right sectors must be put together in a consistent
manner in order to obtain local operators $\bfPhi (z,\zbar)$ with
single-valued correlation functions.  In this section we describe how
to construct such operators that are also $gl(1|1)$ invariant.  

We first need to fix our conventions for the right-moving sector.  
Given the decomposition (\ref{3.3}), we define the right-moving currents
as 
\beq
\label{9.1}
\Hbar = -i \sqrtk  \, \d_\zbar \phibar^1 , ~~~~~
\Jbar = -i \sqrtk \,  \d_\zbar \phibar^2 
\eeq
Right-moving vertex operators of charge 
$(h,j)$ are 
\beq
\label{9.2}
\bar{\CV}^\phi_{h,j} = e^{- i( h \phibar^1 - j \phibar^2 )/\sqrtk}
\eeq
Local $u(1)$ invariant bosonic vertex operators are then
\beq
\label{9.3}
\bfVbos{h}{j} = \Vbos{h}{j} \, \bar{\CV}^\phi_{-h,-j} = 
e^{i( h \bfphi^1 - j \bfphi^2 )/\sqrtk }
\eeq

Imposing locality for the symplectic fermion $\bfchi ( e^{i\alpha} z, 
e^{-i\alpha} \zbar ) = \bfchi (z,\zbar)$ identifies the zero modes
$\chitilde_0 = \bar{\chitilde}_0$ so that the field has the expansion
\beq
\label{9.3b}
\bfchi (z, \zbar) = \bfchi_0 - i \tilde{\bfchi}_0 \log(z\zbar) 
+ i \sum_{n \neq 0} \( 
\inv{n} {\chi_n}  \, z^{-n}  + \inv{n} {\chibar_n} \, \zbar^{-n}  \) 
\eeq
with $\{ \bfchi_0^a , \tilde{\bfchi}_0^b \} = i \ep^{ab} $.  
This implies for example that the local version of the logarithmic
field $\ell (z)$ in (\ref{5.5}) is simply:
\beq
\label{9.4}
\bfell (z , \zbar) = -\inv{2} \ep_{ab} \bfchi^a \bfchi^b 
\eeq
It satisfies
\beq
\label{9.5}
T(z) \bfell (0) \sim \inv{z^2} + \inv{z} \d_z \bfell (0) , ~~~~~
\bar{T} (\zbar) \bfell (0) \sim \inv{\zbar^2} + \inv{\zbar} \d_\zbar \bfell (0) 
\eeq
When we encounter $\chi^1 \chibar^2$ this is thus equated  to $\bfchi^1 \bfchi^2$.  
For the remainder of this section and the next, 
we will not display the local fields $\bfphi, \bfchi, \bfell$ 
in bold face but simply as $\phi, \chi, \ell$. 

\def\bfphi{\phi}
\def\bfchi{\chi}
\def\bfell{\ell}

Let $Q^a =\inv{2\pi i} \oint J^a (z)$ denote the left-moving charge for the 
current $J^a$ and similarly for $\bar{Q}^a$.   The vertex operators 
in the representations  $\rep, \repbar$ satisfy
\beq
\label{9.6}
[Q^a , V^i_\rep] = t^a_{ji}  \, V^j_\rep, ~~~~~
[\bar{Q}^a , \bar{V}^j_\repbar ] = \tbar^a_{ji}  \, \bar{V}^j_\repbar 
\eeq
Introduce the notation
\beq
\label{9.7}
V_\rep \cdot \bar{V}_\repbar = d_{ij} V^i_\rep  \, \bar{V}^j_\repbar
\eeq
The operator $V_\rep \cdot \bar{V}_\repbar$ is invariant under the diagonal
$gl(1|1)$ symmetry $Q^a + \bar{Q}^a$ if the following relation holds:
\beq
\label{9.8}
t^a d + d\,  \tbar^a = 0,  ~~~~~\forall ~ a
\eeq
Using the explicit matrix representations $t^a$ in section IV,  one finds
the following local $gl(1|1)$ invariant operators:

\bigskip

\noindent (i)  {\bf Typical representations with $h\neq j$.~~}    An invariant
is 
\beq
\label{9.9}
\bfPhihj{h}{j} = \V{h}{j} \cdot \bar{V}_{\glrep{1-h}{1-j}} , ~~~~~
d = \( \matrix{0&1\cr -1 & 0 \cr} \)
\eeq
The structure
of the $h,j$ charges is dictated by the $U(1)$ symmetries $H, J$.
It will also prove useful to define
\beq
\label{alter}
\tilde{\bfPhi}_{\glrep{h}{j}} = \bar{V}_{\glrep{1-h}{1-j}}  \cdot \V{h}{j}
\eeq
which can differ from $\bfPhihj{h}{j}$ by  
fermionic signs which arise when left
and right are interchanged.

\bigskip 

\noindent
(ii) {\bf Two-dimensional representations with $h=j$.}   ~~There are four
types of such operators:  

\beq
\label{9.10}
\bfPhihj{h}{h} = \V{h}{h} \cdot \bar{V}_{\glrep{1-h}{1-h} },  ~~~~~
d = \( \matrix{0&1\cr-1&a \, \delta_{h,1/2} \cr } \)
\eeq

\beq
\label{9.11} 
'\bfPhihj{h}{h} ' = V_{\glrep{h}{h}'} \cdot \bar{V}_{\glrep{1-h}{1-h}'} , ~~~~~
d = \( \matrix{ a\,  \delta_{h,1/2} & 1 \cr -1 & 0 \cr } \) 
\eeq

\beq
\label{9.12} 
\bfPhihj{h}{h}' = \V{h}{h} \cdot \bar{V}_{\glrep{1-h}{1-h}'}, ~~~~~
d = \( \matrix{0&0\cr1&0\cr} \) 
\eeq

\beq
\label{9.13} 
'\bfPhihj{h}{h} = V_{\glrep{h}{h}'}  \cdot \bar{V}_{\glrep{1-h}{1-h}}, ~~~~~
d = \( \matrix{0&1\cr0&0\cr} \) 
\eeq
Above $a$ is a free parameter that is only allowed if $h=\inv{2}$ by
$u(1)$ invariance.   

\bigskip

\noindent (iii)  {\bf Four dimensional indecomposable representations.~~}  
Finally there is a local field based on the representation $\hfour{h}$: 
\beq
\label{9.14} 
\bfPhi_{\hfour{h}} = V_{\hfour{h}} \cdot \bar{V}_{\hfour{-h}} , ~~~~~
d = \( \matrix{
0&0&0&1\cr
0&a&-1&0\cr
0&1&0&0\cr
-1&0&0&0\cr } \) 
\eeq
As before, $a$ is a free parameter.

The local fields based on the atypical representations are of interest
since they are expressed in terms of the original local fields
$\bfphi, \bfchi$.   The fields
$\bfPhi'_{\glrep{h}{h}} = k \bfVbos{h-1}{h-1}$ and
$'\bfPhi_{\glrep{h}{h}} = k \bfVbos{h}{h}$ are purely bosonic singlets.  
$\bfPhi_{\glrep{h}{h}}$ and $'\bfPhi'_{\glrep{h}{h}}$ are
fermionic when $a=0$.   Thus  the most interesting field is the logarithmic one 
\beq
\label{9.15}
\bfPhi_{\hfour{h}}  = \bfchi^1 \bfchi^2 \,  \( 
e^{i (h+1) (\phi^1 - \phi^2 )/\sqrtk } + 
e^{i (h-1) (\phi^1 - \phi^2 )/\sqrtk}  \)  + 
a \, k \, 
e^{i h (\phi^1 - \phi^2 ) /\sqrtk }  
\eeq
The case of the $h=0$, which arises in the OPE of
$\bfPhihj{1}{0}$ with $\bfPhihj{0}{1}$, is real:
\beq
\label{9.16}
\bfPhi_{\hfour{0}} = 2 \bfchi^1 \bfchi^2 \,   \cos ((\phi^1 - \phi^2 )/\sqrtk )
\eeq

In the sequel we will need the explicit forms of some additional  local operators
in the case of $k=2$.   The fundamental field involves the twist fields:
\beq
\label{9.17}
\bfPhihj{1}{0} = e^{i \phi^1 /\sqrt{2}}  \, \bfmu_{1/2} + 
e^{i \phi^2/ \sqrt{2} } \,  \bfsigma_{1/2}
\eeq
where $\bfmu_{1/2} = \mu_{1/2} \bar{\mu}_{1/2} $
and $\bfsigma_{1/2} = \sigma^2_{1/2} \bar{\sigma}^1_{1/2} $.  
In addition to the field $\bfPhi_{\hfour{0}}$,  at $k=2$ the
fields $\bfPhihj{2}{0}, \bfPhihj{-1}{1}, \bfPhihj{1}{-1}$ and $\bfPhihj{0}{2}$
are also expressed in terms of the original fields:
\barray
\label{9.18}
\bfPhihj{2}{0} - \tilde{\bfPhi}_{\glrep{-1}{1}} &=& 
4  \d_\mu  \bfchi^1 \d_\mu \bfchi^2  \,  
\cos \(( \bfphi^1 + \bfphi^2)/\sqrt{2} \) 
- 4 \cos( \sqrt{2} \phi^1 ) 
\\ \nonumber
\bfPhihj{1}{-1} - \tilde{\bfPhi}_{\glrep{0}{2}}  
&=&  4\,  \bfchi^1 \bfchi^2  \,  
\cos \( ( \bfphi^1 + \bfphi^2 )/\sqrt{2} \) 
+ 4 (\d_\mu \bfchi^1 \d_\mu \bfchi^2 ) ( \bfchi^1 \bfchi^2 ) \,  
\cos ( \sqrt{2} \bfphi^2 ) 
\earray

\section{Logarithmic perturbations and local lagrangians.}

Using the constructions of the last section,  we can consider a variety of
local perturbations of the free action that preserve $gl(1|1)$.  
The simplest and most interesting are based on the 4-dimensional 
indecomposable representation $\hfour{h}$.  
Consider first a perturbation by  $\bfPhi_{\hfour{0}}$:
\barray
\label{glSG}
S &=& S_{gl(1|1)_k}  +  \int \frac{d^2 x}{8\pi}  ~ \bfPhi_{\hfour{0}} 
\\ \nonumber 
&=&   \int \frac{d^2 x}{8\pi} \(  
 \sum_{a,b=1}^2   \eta_{ab} \,  \d_\mu \bfphi^a \d_\mu \bfphi^b  +
\ep_{ab} \, \d_\mu \bfchi^a \d_\mu \bfchi^b   
~+ g \, 
 \bfchi^1 \bfchi^2 \,  
\cos \( (\bfphi^1 - \bfphi^2 )/\sqrtk  \)  \)
\earray
The above action may be viewed as a $gl(1|1)$ invariant generalization
of the sine-Gordon theory.     The interaction is a $\Delta=0$ logarithmic 
operator.

Next consider  a perturbation by $\bfPhi_{\hfour{1}}$:
\beq
\label{10.1}
S=  \int 
\frac{d^2 x}{8\pi} \( \sum_{a,b=1}^2   \eta_{ab} \,  \d_\mu \bfphi^a \d_\mu \bfphi^b  +
\ep_{ab} \, \d_\mu \bfchi^a \d_\mu \bfchi^b   
+ g \, \bfchi^1 \bfchi^2 \, e^{2i  (\bfphi^1 - \bfphi^2)/ \sqrtk  }
\) 
\eeq
(We set the free parameter $a=0$.)
This may be viewed as a $gl(1|1)$ invariant Liouville theory. As for the usual
Liouville, 
background charges $\vec{q}_0$ can be introduced to give the perturbation
$\Delta = 1$:
\beq
\label{10.2bc}
T_{\phivec}  = -\inv{2}  \d \phivec \cdot \d \phivec  + \frac{i}{2} \qvec_0 \cdot \d^2 \phivec 
\eeq
where $\phivec\cdot \phivec \equiv 
 \eta_{ab} \phi^a \phi^b$.  The dimensions are
\beq
\label{10.3bc}
\Delta(e^{i \qvec \cdot \phivec} ) = \inv{2} \qvec \cdot ( \qvec - \qvec_0 )
\eeq
The new central charge in the bosonic sector is $c_{\rm bosonic} = 2 - 3 \qvec_0 \cdot \qvec_0 $.
 For the perturbation in (\ref{10.1}), $\qvec = 2 (1,1)/\sqrtk$ with $\qvec\cdot \qvec = 0$.  
 Choosing $\qvec_0 = \sqrtk ( -1,  1)/2 $ endows it with dimension $1$.  Note that
 since $\qvec_0 \cdot \qvec_0 = 0$,  the total central charge remains zero.  
Note also  that $\qvec_0 = \sqrtk ( -1, 1)$ renders the full cosine term of
the sine-Gordon version  with $\Delta = 1$.    We will not pursue adding 
background charges further in this paper.

An important feature of logarithmic perturbations such as in eq. (\ref{glSG}) 
is the following.   Because of the indefinite metric for the bosons,
OPE's of the dimension zero operators in the bosonic sector are   regular:
\beq
\label{regu}
e^{ih( \phi^1 - \phi^2 )(z)}  \,  e^{i h' (\phi^1 - \phi^2 )(w)}     \sim {\rm regular} 
\eeq
Therefore in perturbation theory,  the perturbation by $\bfPhi_{\hfour{0}}$ 
behaves like a mass term $\chi^1 \chi^2$.      As for a mass term,  it simply leads to logarithmic
corrections to correlation functions without changing the exponents.  

More generally consider a conformal field theory perturbed by a 
logarithmic operator $\Phi_\ell$  with action
\beq
\label{log.1}
S = S_{\rm cft}  +  \int  \frac{d^2 x}{2\pi}  ~  g  \,  \Phi_\ell (x)
\eeq
The RG beta function for $g$ is determined by OPE of $\Phi_\ell$ with itself.
If $\Phi_\ell$  has $\Delta =0$,  then the singular term in the OPE is
at worse a logarithm: 
\beq
\label{log.2}
\Phi_\ell (x) \,  \Phi_{\ell} (0)  =  \gamma \,  \log (x^2) \,  \Phi_{\ell} (0) + ...
\eeq
Introducing a short distance cut-off $a$,  
$\int_a d^2 x \, \log (x^2) = \pi a^2 ( 1- \log (a)) + {\rm const.} $,  
then the cut-off dependent coupling is $g (a) = g + 
\gamma g^2  a^2 (1-2\log (a))/4$.    This implies that
as $a\to 0$,  the beta function $d g(a) / d \log (a)  \to 0$.   
Thus, in general one does not expect logarithmic perturbations to
drive the theory to a new fixed point.   
This feature was also discussed in 
\cite{cauxlog}.

\section{Aspects of $\osphat_{-2}$}

Normally,  larger dimensional algebras such as $\osphat_k$ require more
fields to be represented.  However it was shown by Ludwig\cite{LudwigTwist} that
the  special case of $k=-2$ has a free
field representation with the same field content as above.      In this
section we explain how this follows from our results on  $\glk$ and
use this connection to present additional results.

First note that $\osphat_{-2}$ has two $\glhat_{-2}$ subalgebras 
which do not commute.  Let us try to represent $H,J$ and both $S_\pm$
and $\Shat_\pm$ with the same field content as in eq. (\ref{3.5}).    The problem
with generic $k$ is that the OPE 
$~S_+ (z) \Shat_+ (0) \propto  1/(z^{2+ 2/k})$ and thus  does not close on integer powers,  except
for $k=-2$.    The resulting free field representation at $k=-2$ then follows
from previous expressions (\ref{3.5}) with $\sqrtk = i \sqrt{2}$:
\barray
\label{10.1b}
H &=& - \sqrt{2}  \, \d\phi^1 , ~~~~J= -\sqrttwo \, \d \phi^2 , ~~~~~
J_\pm = \pm 2\,  e^{\mp \sqrttwo \phi^2 }
\\ \nonumber
S_\pm &=& \pm i\sqrt{2}  ~ \d \chi^\pm ~ e^{\pm (\phi^1 - \phi^2 )/\sqrttwo} 
, 
~~~~~
\Shat_\pm = \pm i \sqrttwo ~ \d \chi^\mp  ~
 e^{ \mp (\phi^1 + \phi^2 )/\sqrttwo}  
\earray 
where here for notational simplicity 
we have defined $\bfchi^{1,2}  = \bfchi^{+,-}$. 
The OPE's of the above currents is the same as in Appendix A up to some
inconsequential 
minus signs. 
Note that $J, J_\pm$ have  the standard $su(2)_1$ representation
in terms of a single boson.  
For the remainder of this section $\glhat$ refers to the  
$\glhat_{-2}$  algebra
generated by $H, J, S_\pm$.

The finite dimensional representations of $osp(2|2)$ can be labeled by
the $su(2)$ with generators $J, J_\pm $ and by the $u(1)$ charge $H$.   
The typical, irreducible representations will be denoted as 
$\osprep{b}{s}$ where $s\in \{ 0, \inv{2} , 1, \frac{3}{2},....\}$ 
is an $su(2)$ spin and $b=H/2$.  
These representations are $8s$ dimensional.   In order to describe their
$su(2) \otimes u(1)$ decomposition,  let $\surep{b}{s}$ denote the
$2s+1$ dimensional representation with $J/2  = s_3 = -s, -s+1, ..., s$
and $H=2b$.    The generic decomposition is\cite{osp,Serban} 
\beq
\label{10.2}
\osprep{b}{s} = \surep{b}{s} \oplus \surep{b+\texthalf}{s-\texthalf}  
\oplus  \surep{b-\texthalf}{s-\texthalf} \oplus \surep{b}{s-1} 
\eeq

The stress tensor is built from the single quadratic
casimir\cite{Serban}:
\beq
\label{ospstress}
T_{osp(2|2)} = \inv{ 2(2-k) } 
\[ 
J^2 - H^2 - \inv{2} (J_+ J_- + J_- J_+ ) 
+ (S_+ S_- - S_- S_+ ) + (\Shat_- \Shat_+ - \Shat_+ \Shat_- ) 
\]
\eeq
Since the $k=-2$ case has the same free field construction as $\glhat_{-2}$,
and there exists the $k\to -k$ automorphism (\ref{auto}),  one
must have $T_{\osphat_{-2}} = T_{\glhat_{-2}} = T_{\glhat_{2}}$. 
The typical representations with $b^2 \neq s^2$ have conformal
dimension
\beq
\label{10.3}
\Delta^{osp}_{[b,s]} =  \frac{ 2 (s^2 - b^2 )}{2-k} 
\eeq

The vertex operators follow from the results of section VI and the
decomposition of the $osp(2|2)$ representations in terms of $gl(1|1)$.
The later can be deduced from (\ref{10.2}) with the identification
$H=2b, J = 2 s_3$.   For example
\beq
\label{10.4}
\osprep{b}{\smallhalf} = \glrep{2b}{1} \oplus \glrep{2b+1}{0}
 \eeq
 where $\glrep{h}{j}$ are the 2-dimensional $gl(1|1)$ representations of
 section IV.   A check of the above is the scaling dimension:
 \beq
 \label{10.5}
 \Delta^{osp}_{[b,\inv{2}]} = 
\Delta^{gl}_{\glrep{2b}{1}} =  \Delta^{gl}_{\glrep{2b+1}{0}} 
 \eeq
 where $\Delta^{gl}_{\glrep{h}{j}}$ are the  $\glhat$ scaling dimensions in eq. (\ref{4.9}) 
 at $k=-2$.  
 
 The vertex operator for the 4 dimensional representation $\osprep{b}{\inv{2}}$ is then
 \beq
 \label{10.6}
 V^{osp}_{[b, \inv{2}]} = \( \matrix{
 V_{\glrep{2b}{1}} \cr
 V_{\glrep{2b+1}{0} }\cr } 
 \) 
 \eeq
 Throughout this section $V_{\glrep{h}{j}}$ refers to the $k=-2$ vertex operators,
 which are simply related to the  $k=2$ expressions in section VI  by
 the automorphism (\ref{auto}).

As for $gl(1|1)$, there are atypical, indecomposable but reducible
representations at $b^2 = s^2$.   The simplest is $8$ dimensional 
and arises in the following tensor product
\beq
\label{10.7}
\osprep{0}{\texthalf} \otimes \osprep{0}{\texthalf} = 
\osprep{0}{1} \oplus \eight 
\eeq
The $gl(1|1)$ decomposition is
\beq
\label{10.7b} 
\eight = \hfour{0} + \glrep{2}{0} + \glrep{-1}{1} 
\eeq
and the vertex operator is
\beq
\label{10.8}
V^{osp}_{[8]} =  \( \matrix{
\V{2}{0} \cr
V_{\hfour{0}} \cr
\V{-1}{1} \cr } \) 
\eeq
This is a $\Delta =0$ logarithmic operator due to the presence of 
$V_{\hfour{0}}$,  and reflects the fact that the quadratic casimir is
not diagonal on  the $\eight$-representation\cite{Serban}.    The structure of
the $\eight$ is shown in Figure 2.    The explicit form of the
$\V{2}{0}, \V{-1}{1}$ operators in (\ref{10.8})  can found by
acting on $V_{\hfour{0}}$ with the generators according to
Figure 2,  where $V_{\hfour{0}}$ is given in eq. (\ref{6.9}) 
with $k=-2$;  the result is consistent with eq. (\ref{7.10}).  

\begin{figure}[htb] 
\begin{center}
\hspace{-15mm}
\includegraphics[width=10cm]{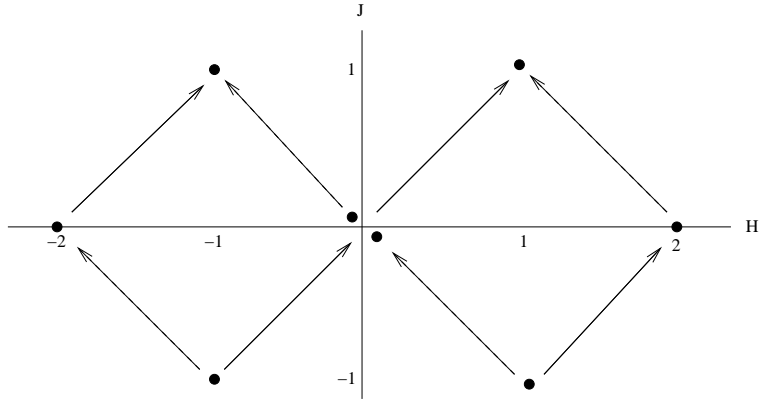} 
\end{center}
\caption{The structure of the $osp(2|2)$ representation $\eight$. 
The NE  arrows (to the right) indicate the action of $S_+$ and the
dashed NW arrows indicate the action of $\Shat^+$.} 
\vspace{-2mm}
\label{Figure2} 
\end{figure}

In the next section we will need the local operator $\bfPhi_{[8]}^{osp} 
= V^{osp}_{[8]} \cdot \bar{V}^{osp}_{[8]}$, which can
be constructed as in section IV.   Setting $k=-2$ and recalling
the automorphism (\ref{auto}), one finds
\beq
\label{10.9}
\bfPhi_{[8]}^{osp} = 
 \bfPhi_{\glrep{2}{0}} + \bfPhi_{\glrep{-1}{1} } + 2 \bfPhi_{\hfour{0}}
\eeq
The explicit form is
\beq 
\label{10.10}
\bfPhi_{[8]}^{osp} =  4\,  \bfchi^1 \bfchi^2   \(  
\cosh(  ( \bfphi^1 - \bfphi^2 )/\sqrttwo )
 + \cosh(( \bfphi^1 + \bfphi^2 )/\sqrttwo )  \) 
+ 4  (\d_\mu \bfchi^1 \d_\mu \bfchi^2 ) ( \bfchi^1 \bfchi^2 ) ~ 
\cosh(\sqrttwo \bfphi^1) 
\eeq
(As in section IX, above the fields $\phi, \chi$ refer to the
local fields $\pmb{$\phi$} (z, \zbar)$ and $\pmb{$\chi$} (z, \zbar)$.)

\section{Application to the spin quantum Hall transition.}

In this section we apply some of the tools developed so far
to the spin quantum Hall transition (SQHT).  
Since this is tangential to the original scope of this article,
details and additional results will be presented elsewhere\cite{SuperDisorder}. 
 Let us
begin with a short summary of the relevant background. 
   Like the usual
quantum Hall transition, the SQHT has a network model 
description\cite{Chalker} and can be mapped onto a spin
chain\cite{Senthil}.   
Gruzberg et. al. mapped the spin chain onto percolation.
Critical percolation explains the two main exponents 
that were studied numerically\cite{Senthil},  namely the correlation
length $\nu_{\rm perc.} = 4/3$ and the density of states
exponent $\rho (E) \sim E^{1/7}$.

The SQHT  can also be formulated in the continuum as
a model of disordered Dirac fermions.   It has
an $su(2)$ gauge disorder with coupling $g_s$, and
two additional kinds of mass/potential disorder 
with couplings $g_c$ and $g_8$.  
(See \cite{SpinCharge} 
for precise definitions.)  
  Whereas $g_s$ 
corresponds to current-current interactions for the
$su(2)_{k=0}$ currents,   the coupling $g_c$ 
corresponds to  the 
$\osphat_{-2}$ currents.         The renormalization group
flow for the couplings was studied in \cite{SpinCharge,networkRG}.  
Though a perturbative fixed point was not found 
in the latter work,  one feature related to the super spin-charge
separation described in section VIII emerged as follows.  
If the $g_8$ disorder is initially set to zero,  
which amounts to an initial  fine-tuning of the model, then 
the RG flow of the remaining couplings decouples due
to the spin charge separation.   At one-loop,  $d g_s / d \ell = g_s^2$
and $d g_c / d \ell = - 2 g_c^2$, and this decoupling persists to
higher orders.    As described more generally in section VIII,
 whereas $g_s$ is marginally relevant, 
$g_c$ is marginally irrelevant,  so that the fixed point 
of the model at $g_8=0$ was argued to be $\osphat_{-2}$\cite{SpinCharge}.  
For another approach based on replicas, see \cite{FendleyKonik}. 

The fixed point $osp(2|2)_{-2}$ reproduces the main  exponents
of the SQHT.   This is most transparent using the 
$gl(1|1)_2$ embedding, since, as explained in section VII, 
it has precisely the percolation exponents that 
are relevant for the SQHT.    In the $osp(2|2)_{-2}$ 
description,  the density operator $\rho$ is
identified with the representation $[0,\inv{2}]^{osp}$  with
$\Delta = \inv{8}$,  and determines the density of states
exponent, $\inv{7}  = \frac{\Delta}{1-\Delta}$.  
The $\Delta = \frac{5}{8}$ field which determines $\nu_{\rm perc.} = 4/3$
is a descendant of the $\Delta = -\frac{3}{8}$ field 
$[\pm 1,\inv{2}]^{osp}$.    Note also that $1$-hull operator
with $\Delta = \inv{3}$ in the theory of percolation is not
contained in the $gl(1|1)_2$ theory. (See section VII.)  
This appears to be consistent with the fact this operator
does not play any known r\^ole in the 
SQHT.  Related comments were made in \cite{ReadSaleur},
where there also the $\Delta= \inv{3}$ field was not in
the spectrum.  

The potential  problem with the $\osphat_{-2}$ fixed point is that
the residual  $g_8$ perturbations potentially modify it. 
  This is consistent
with the study of the spectrum of the spin chain in \cite{ReadSaleur}
which suggested that the critical  point is a new kind of theory
with $osp(2|2)$ symmetry that  is not simply  a current algebra.    
The higher-order corrections to the beta functions computed in
\cite{networkRG}, which are 
 correct up to at least 4-loops\cite{Ludwig4loop},
do not help to  resolve the problem since the flow is to a  singular point. 

To resolve these difficulties,  we propose to carry out
the RG flow in two stages.   First one sets $g_8 =0$ and flows
to $\osphat_{-2}$.  In the second stage we restore the $g_8$
coupling as a perturbation of the current algebra.  
The currents in the $g_8$ coupling transform under the $\eight$ of
$osp(2|2)$ and the spin $1$ of the $su(2)$.  In the RG flow the
$su(2)$ is gapped out which leaves a field transforming
under the $\eight$.   The resulting action is
\beq
\label{11.1}
S = S_\free + g_8 \, \int \frac{d^2 x}{2 \pi} ~  \bfPhi_{[8]}^{osp} (x)
\eeq
where $S_\free$ is just the free action for the scalars and
symplectic fermion eq. (\ref{3.1}) and $\bfPhi_{[8]}^{osp}$ is the
logarithmic operator in eq. (\ref{10.10}).

The proposal eq. (\ref{11.1}) overcomes previous difficulties
in a number of ways.  First, as argued  in section X,
the dimension zero logarithmic perturbation by $\bfPhi_{[8]}^{osp}$
does not modify the scaling dimensions but only leads to
logarithmic corrections to the correlation functions.   
Second,  the model has an $osp(2|2)$ symmetry, as expected
from the spin-chain description.  However,  because of the
logarithmic perturbation,  the critical point is not strictly speaking
a conformal current algebra, even though it has the same
exponents as the current algebra.   This is consistent
with observations made in \cite{ReadSaleur}.

Further checks of this proposal will be described in \cite{SuperDisorder},
where we explain  how to obtain the multi-fractal exponents.

\section{Conclusions}

To summarize,  using the detailed properties of
the twist and logarithmic fields in the symplectic fermion
sector, we have  explicitly constructed all the 
primary fields of the $gl(1|1)_k$ current algebra
at arbitrary level $k$.  We have also identified
a closed operator algebra at integer level.   
For the indecomposible representations,  the  explicit
construction of the logarithmic operators led to 
$gl(1|1)$ invariant models as perturbations by these operators,
and the simplest have local lagrangians that generalize
the Liouville and sine-Gordon models.    We also argued 
that these logarithmic perturbations have a trivial beta functions
and just give logarithmic corrections to the correlation functions
without changing the anomalous dimensions.   
We derived a new form of super  spin charge separation and gave
general arguments indicating how the $gl(1|1)_N$ theory
can arise as a critical point of disordered Dirac fermions
in $2+1$ dimensions.   By studying the $gl(1|1)$ embeddings,
we also constructed explicitly the local logarithmic field corresponding
to the indecomposable representation $\eight$ of $osp(2|2)_{-2}$. 
Since other super current algebras typically have
$gl(1|1)_k$ subalgebras, it should be possible to
obtain other new results as well.   

We initiated the application of these new tools to the investigation
of critical points of disordered Dirac fermions
by re-examining the spin quantum Hall transition. 
It was shown that the 1-copy theory is a perturbation of
the $osp(2|2)_{-2}$ current algebra by the logarithmic
field corresponding to the $\eight$ indecomposable representation.  
In \cite{SuperDisorder} we will extend this analysis to
$N$-copies and thereby compute the  multi-fractal exponents.  
We will also apply these methods to the original Chalker-Coddington network model
for the ordinary quantum Hall transition,  where
we essentially obtain the $gl(1|1)$ invariant sine-Gordon model
eq. (\ref{glSG}).     These
results are not presented here since they require 
more specific details about the disordered Dirac
fermion theories.

\section{Acknowledgments}

I would like to thank the organizers of the program {\it Strong fields, Integrability, and Strings}
at the Isaac  Newton Institute for Mathematical Sciences  during which this work was
begun in July 2007.

\section{Appendix A: complete $\ospk$ relations}

\def\Sh{\Shat}

Our conventions for the $osp(2|2)_k$ current algebra are  based
on the level $1$ representation in terms of $\psi_\pm , \beta_\pm$ given
in section II:

\begin{eqnarray}
\nonumber
J(z) J(0) &\sim& - \frac{k}{z^2} ,  
~~~~~~~~~~~~~~
~H(z) H(0) \sim \frac{k}{z^2}
\\ \nonumber
J(z) J_\pm (0) &\sim&  \pm \frac{2}{z} ~ J_\pm
, ~~~~~~~~~~
J_+(z) J_- (0) \sim \frac{2k}{z^2} - \frac{4}{z} J
\\ \nonumber
J(z) S_\pm (0) &\sim&  \pm \inv{z} S_\pm   ,  
~~~~~~~~~~ J(z) \Sh_\pm (0) \sim
\pm \inv{z} \Sh_\pm
\\ \nonumber
H(z) S_\pm (0) &\sim&  \pm \inv{z} S_\pm  , ~~~~~~~~~~
H(z) \Sh_\pm (0) \sim \mp \inv{z} \Sh_\pm
\\ \label{3.8}
J_\pm (z) S_\mp (0) &\sim& \frac{2}{z} \Sh_\pm , 
~~~~~~~~~~~~~~~
J_\pm (z) \Sh_\mp (0) \sim - \frac{2}{z} S_\pm
\\
\nonumber
&&
S_\pm (z) \Sh_\pm (0) \sim \pm \inv{z} J_\pm
\\ \nonumber
&&
S_+ (z) S_- (0) \sim  \frac{k}{z^2} + \inv{z} (H-J)
\\ \nonumber
&&
\Sh_+ (z) \Sh_- (0) \sim - \frac{k}{z^2} + \inv{z} (H+ J)
\end{eqnarray}

\section{Appendix B:  More on symplectic fermions}

In this appendix we provide some derivations of the results used in this paper.
Most are already contained in \cite{FMS, Kausch}.    

First consider the first-order system for the bosonic $\beta_\pm$ ghosts eq. (\ref{2.1})
with $c=-1$.    It was shown in \cite{FMS}   
 that these can be bosonized in terms of a single scalar field $\phi$ for the $U(1)$ current
 and an additional auxiliary fermionic $\eta-\xi$ system with $c=-2$:
 \beq
 \beta_+ = e^{i\phi}  \, \eta  , ~~~~~\beta_- = e^{- i \phi} \d \xi 
 \label{B.1}
 \eeq
 where $\langle \phi (z) \phi (0) \rangle =\log (z)$.    
The $\eta-\xi$ system is also first order,
 with action
 \beq
 \label{B.2}
 S_{\eta, \xi}
 = \inv{4\pi} \int d^2 x \(  \eta \d_\zbar \xi + \bar{\eta} \d_z \bar{\xi}  \)
 \eeq
 but now with $\Delta (\eta, \xi) = (1,0)$.  
 
 Before using the equations of motion,  we can relate the model to 
 symplectic fermions by identifying
 $\eta = i \d_z \bfchi^1, ~\bar{\eta} = i \d_\zbar \bfchi^1 $, 
 $\d_\zbar \xi = i \d_\zbar \bfchi^2 ,  \d_z \bar{\xi} = i \d_z \bfchi^2$ .
 In this way one  obtains the  second-order action (\ref{3.1}).    
 After using the equations of motion, the chiral components 
 are identified as follows:
 \beq
 \label{B.3}
 \eta (z) = i \d_z \chi^1 , ~~~~~ \xi (z)  = i \chi^2 (z) 
 \eeq
 consistent with the conformal dimensions.   
 It is important to note that the $\eta-\xi$ system does not contain
 the zero mode of $\chi^1$,  and thus does not  explicitly contain the
 logarithmic operator $\ell$ in eq. (\ref{3.5}).   
 
 The $\eta-\xi$ system can in turn be bosonized with a single scalar $f(z)$ 
 \beq
 \label{B.4}
 \eta = e^{-i f} , ~~~~~\xi = e^{if} 
 \eeq
 with $\langle f(z) f(0) \rangle = - \log (z)$.   However, in order to obtain $\Delta (\eta, \xi ) = (1,0)$, 
 $f$ has a background charge:
 \beq
 \label{B.5}
 T_f = - \inv{2} (\d f)^2 + \frac{i}{2} \d^2 f
 \eeq
 With this background charge,
 \beq
 \label{B.5b}
 \Delta (e^{i\alpha f}) = \frac{\alpha (\alpha -1)}{2}  
 \eeq
 and the correlation functions have the charge asymmetry:
 \beq
 \label{B.6}
 \langle e^{i \lambda f} \,  e^{i \lambda' f} \rangle  \neq 0  ~~~~~~\iff \lambda + \lambda' =1 
 \eeq
 All the correlation functions can be computed with Coulomb gas techniques.  
 
 \def\sector#1{ [#1] }
 
 Let $\sector\lambda$ denote the sector $e^{i \lambda f} |0\rangle$ and its decendents.  
 The twist field $\mu_\lambda \in \sector\lambda$.   Furthermore since
 $\chi^1 \in \sector{-1}, \chi^2 \in \sector{1}$,  then $\sigma^1_\lambda \in \sector{\lambda -1}$
 and $\sigma^2_\lambda \in \sector{\lambda +1}$.   In this way, one obtains
 the conformal dimensions eqs. (\ref{5.11}, \ref{5.12}).  
 
 The OPE's (\ref{5.10})  are derived as follows.  Define
 $ | \mu_\lambda \rangle = \mu_\lambda (0) | 0 \rangle$ and
 $\langle  \mu_{1-\lambda} |  = \lim_{z \to \infty} z^{2 \Delta_\lambda}  \,
 \langle 0 | \mu_{1-\lambda} (z) $.   Using the mode expansions (\ref{5.9}) 
 and $\chi^a_{n-\lambda} | \mu_\lambda \rangle = 0$ for $n>0$, 
 \beq
 \label{B.7}
 \langle \mu_{1-\lambda} | \, \d_z \chi^1 (z) \,   \chi^2 (w)  \, 
 | \mu_\lambda \rangle = - \( \frac{w}{z} \)^\lambda \inv{z-w} 
 \eeq
 Taking the derivative $\d_w$ of the above equation,  letting $w\to 0$ 
 and $z \to \infty$,  and using $\langle \mu_{1-\lambda} | \mu_\lambda \rangle = 1$, 
 one obtains the $\sqrt{\lambda}$ factor in eq. (\ref{5.10}).    
 The $\sqrt{1-\lambda}$ factors are obtained similarly.

\vspace{.4cm}
\end{document}